\documentclass[11pt]{article}
\author{Gabriel Nivasch\thanks{\texttt{gabriel.nivasch@cs.tau.ac.il}.
Work was supported by ISF Grant 155/05 and by the Hermann
Minkowski--MINERVA Center for Geometry at Tel Aviv University.}
\quad Micha Sharir\thanks{\texttt{michas@post.tau.ac.il}. Work was
partially supported by NSF grant CCF-05-14079, by a grant from the
U.S.-Israel Binational Science Foundation, by ISF Grant 155/05, and
by the Hermann Minkowski--MINERVA Center for Geometry
at Tel Aviv University.}\\
\footnotesize School of Computer Science\\[-1mm]
\footnotesize Tel Aviv University\\[-1mm]
\footnotesize Tel Aviv 69978, Israel}
\date{July 27, 2008}

\title{Eppstein's bound on intersecting triangles revisited}

\usepackage{amsmath,amsfonts,amssymb,amsthm}
\usepackage{hyperref}
\usepackage{graphics}

\newtheorem{lemma}{Lemma}

\begin{document}

\maketitle

\begin{abstract}
Let $S$ be a set of $n$ points in the plane, and let $T$ be a set of
$m$ triangles with vertices in $S$. Then there exists a point in the
plane contained in $\Omega(m^3/(n^6\log^2 n))$ triangles of $T$.
Eppstein (1993) gave a proof of this claim, but there is a problem
with his proof. Here we provide a correct proof by slightly
modifying Eppstein's argument.

\emph{Keywords:} Triangle; Simplex; Selection Lemma; $k$-Set
\end{abstract}

\section{Introduction}

Let $S$ be a set of $n$ points in the plane in general position (no
three points on a line), and let $T$ be a set of $m \le {n\choose
3}$ triangles with vertices in $S$. Aronov et al.~\cite{ACEGSW}
showed that there always exists a point in the plane contained in
the interior of
\begin{equation}\label{eq_bound_log5}
\Omega{\left({ m^3 \over n^6\log^5 n }\right)}
\end{equation}
triangles of $T$. Eppstein \cite{eppstein} subsequently claimed to
have improved this bound to
\begin{equation}\label{eq_bound_log2}
\Omega{\left({ m^3 \over n^6\log^2 n }\right)}.
\end{equation}
There is a problem in Eppstein's proof, however.\footnote{The very
last sentence in the proof of Theorem 4 (Section 4) in
\cite{eppstein} reads: ``So $\epsilon = 1/2^{i+1}$, and $x =
m\epsilon/y = O(m/8^i)$, from which it follows that $x/\epsilon^3 =
O(n^2)$.'' This is patently false, since what actually follows is
that $x/\epsilon^3 = O(m)$, and the entire argument falls through.}
In this note we provide a correct proof of (\ref{eq_bound_log2}), by
slightly modifying Eppstein's argument.

\subsection{The Second Selection Lemma and $k$-sets}

The above result is the special case $d=2$ of the following lemma
(called the \emph{Second Selection Lemma} in \cite{matou}), whose
proof was put together by B\'ar\'any et al.~\cite{BFL}, Alon et
al.~\cite{ABFK}, and \v Zivaljevi\'c and Vre\'cica \cite{ZV}:

\begin{lemma}\label{lemma_2nd_sel}
If $S$ is an $n$-point set in $\mathbb R^d$ and $T$ is a family of
$m\le {n \choose d+1}$ $d$-simplices spanned by $S$, then there
exists a point $p\in \mathbb R^d$ contained in at least
\begin{equation}\label{eq_2nd_SL}
c_d \left({ m \over n^{d+1}}\right)^{s_d} n^{d+1}
\end{equation}
simplices of $T$, for some constants $c_d$ and $s_d$ that depend
only on $d$.
\end{lemma}

(Note that $m/n^{d+1} = O(1)$, so the smaller the constant $s_d$,
the stronger the bound.) Thus, for $d=2$ the constant $s_2$ in
(\ref{eq_2nd_SL}) can be taken arbitrarily close to $3$. The general
proof of Lemma \ref{lemma_2nd_sel} gives very large bounds for
$s_d$; roughly $s_d \approx (4d+1)^{d+1}$.

The main motivation for the Second Selection Lemma is deriving upper
bounds for the maximum number of \emph{$k$-sets} of an $n$-point set
in $\mathbb R^d$; see \cite[ch.~11]{matou} for the definition and
details.

\section{The proof}

We assume that $m = \Omega( n^2 \log^{2/3} n)$, since otherwise the
bound (\ref{eq_bound_log2}) is trivial. The proof, like the proof of
the previous bound (\ref{eq_bound_log5}), relies on the following
two one-dimensional \emph{selection lemmas} \cite{ACEGSW}:

\begin{lemma}[Unweighted Selection Lemma]\label{lemma_unw}
Let $V$ be a set of $n$ points on the real line, and let $E$ be a
set of $m$ distinct intervals with endpoints in $V$. Then there
exists a point $x$ lying in the interior of $\Omega(m^2/n^2)$
intervals of $E$.
\end{lemma}

\begin{lemma}[Weighted Selection Lemma]\label{lemma_weighted}
Let $V$ be a set of $n$ points on the real line, and let $E$ be a
\emph{multiset} of $m$ intervals with endpoints in $V$. Then there
exists a multiset $E'\subseteq E$ of $m'$ intervals, having as
endpoints a subset $V' \subseteq V$ of $n'$ points, such that all
the intervals of $E'$ contain a common point $x$ in their interior,
and such that
\begin{equation*}
{m'\over n'}  = \Omega{\left({m\over n\log n}\right)}.
\end{equation*}
\end{lemma}

The proof of the desired bound (\ref{eq_bound_log2}) proceeds as
follows:

Assume without loss of generality that no two points of $S$ have the
same $x$-coordinate. For each triangle in $T$ define its \emph{base}
to be the edge with the longest $x$-projection. For each pair of
points $a, b\in S$, let $T_{ab}$ be the set of triangles in $T$ that
have $ab$ as base, and let $m_{ab} = |T_{ab}|$. (Thus, $\sum_{ab}
m_{ab} = m$.)

Discard all sets $T_{ab}$ for which $m_{ab} < m/n^2$. We discarded
at most ${n\choose 2} m/n^2 < m/2$ triangles, so we are left with a
subset $T'$ of at least $m/2$ triangles, such that either $m_{ab} =
0$ or $m_{ab} \ge m/n^2$ for each base $ab$.\footnote{This critical
discarding step is missing in \cite{eppstein}, and that is why the
proof there does not work.}

Partition the bases into a logarithmic number of subsets $E_1, E_2,
\ldots, E_k$ for $k = \log_4 (n^3/m)$, so that each $E_j$ contains
all the bases $ab$ for which
\begin{equation}\label{eq_bound_m_ab}
{4^{j-1} m \over n^2} \le m_{ab} < {4^j m \over n^2}.
\end{equation}
Let $T_j = \bigcup_{ab \in E_j} T_{ab}$ denote the set of triangles
with bases in $E_j$, and $m_j = |T_j|$ denote their number. There
must exist an index $j$ for which
\begin{equation*}
m_j \ge 2^{-(j+1)} m,
\end{equation*}
since otherwise the total number of triangles in $T'$ would be less
than $m/2$. From now on we fix this $j$, and work only with the
bases in $E_j$ and the triangles in $T_j$.

For each pair of triangles $abc$, $abd$ having the same base $ab \in
E_j$, project the segment $cd$ into the $x$-axis, obtaining segment
$c'd'$. We thus obtain a multiset $M_0$ of horizontal segments, with
\begin{equation*}
|M_0| \ge {m_j\over 2} \left( {4^{j-1} m \over n^2} - 1 \right) =
\Omega{\left({ 2^j m^2 \over n^2 }\right)}.
\end{equation*}
(Each of the $m_j$ triangles in $T_j$ is paired with all other
triangles sharing the same base, and each such pair is counted
twice.)

We now apply the Weighted Selection Lemma
(Lemma~\ref{lemma_weighted}) to $M_0$, obtaining a multiset $M_1$ of
segments delimited by $n_1$ distinct endpoints, all segments
containing some point $z_0$ in their interior, with
\begin{equation*}
{ |M_1| \over n_1} = \Omega{\left({ |M_0| \over n \log n }\right)} =
\Omega{\left({ 2^j m^2 \over n^3 \log n }\right)}.
\end{equation*}

\begin{figure}
\centerline{\includegraphics{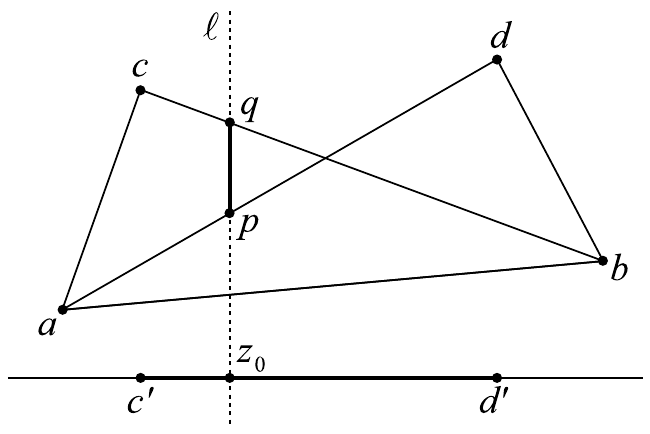}}
\caption{\label{fig_triangle_pair} Pairing two triangles with a
common base.}
\end{figure}

Let $\ell$ be the vertical line passing through $z_0$. For each
horizontal segment $c'd' \in M_1$, each of its (possibly multiple)
instances in $M_1$ originates from a pair of triangles $abc$, $abd$,
where points $a$ and $c$ lie to the left of $\ell$, and points $b$
and $d$ lie to the right of $\ell$. Let $p$ be the intersection of
$\ell$ with $ad$, and let $q$ be the intersection of $\ell$ with
$bc$. Then, $pq$ is a vertical segment along $\ell$, contained in
the union of the triangles $abc$, $abd$ (see Figure
\ref{fig_triangle_pair}). Let $M_2$ be the set of all these segments
$pq$ for all $c'd' \in M_1$.

Note that the vertical segments in $M_2$ are all distinct, since
each such segment $pq$ uniquely determines the originating points
$a$, $b$, $c$, $d$ (assuming $z_0$ was chosen in general position).

Let $n_2$ be the number of endpoints of the segments in $M_2$. We
have $n_2 \le n n_1$, since each endpoint (such as $p$) is uniquely
determined by one of $n_1$ ``inner'' vertices (such as $d$) and one
of at most $n$ ``outer'' vertices (such as $a$).

Next, apply the Unweighted Selection Lemma (Lemma~\ref{lemma_unw})
to $M_2$, obtaining a point $x_0\in \ell$ that is contained in
\begin{equation*}
\Omega{\left({ |M_2| ^2 \over n_2^2 }\right)} = \Omega{\left(
{1\over n^2} \left({ |M_1| \over n_1 }\right)^2 \right)} =
\Omega{\left({ 4^j m^4 \over n^8 \log^2 n }\right)}
\end{equation*}
segments in $M_2$. Thus, $x_0$ is contained in at least these many
\emph{unions of pairs of triangles} of $T_j$. But by
(\ref{eq_bound_m_ab}), each triangle in $T_j$ participates in at
most $4^j m / n^2$ pairs. Therefore, $x_0$ is contained in
\begin{equation*}
\Omega{\left({ m^3 \over n^6 \log^2 n }\right)}
\end{equation*}
triangles of $T_j$.

\section{Discussion}

Eppstein \cite{eppstein} also showed that there always exists a
point in $\mathbb R^2$ contained in $\Omega(m/n)$ triangles of $T$.
This latter bound is stronger than (\ref{eq_bound_log2}) for small
$m$, namely for $m = O(n^{5/2} \log n)$.

On the other hand, as Eppstein also showed \cite{eppstein}, for
every $n$-point set $S$ in general position and every $m =
\Omega(n^2)$, $m\le {n\choose 3}$, there exists a set $T$ of $m$
triangles with vertices in $S$, such that no point in the plane is
contained in more than $O(m^2/n^3)$ triangles of $T$. Thus, with the
current lack of any better lower bound, the bound
(\ref{eq_bound_log2}) appears to be far from tight. Even achieving a
lower bound of $\Omega(m^3/n^6)$, without any logarithmic factors,
is a major challenge still unresolved.

It is known, however, that if $S$ is a set of $n$ points in $\mathbb
R^3$ in general position (no four points on a plane), and $T$ is a
set of $m$ triangles spanned by $S$, then there exists a \emph{line}
(in fact, a line spanned by two points of $S$) that intersects the
interior of $\Omega(m^3/n^6)$ triangles of $T$; see \cite{DE} and
\cite{smo_phd} for two different proofs of this.

\end{document}